\documentclass[useAMS,usenatbib]{mn2e}
\usepackage{graphicx}

\title[Physical properties of Seyfert nuclei with EVN]{Physical properties of the nuclear region in Seyfert galaxies derived from observations with the European VLBI Network}
\author[Bontempi, Giroletti, Panessa, Orienti, Doi]{P. Bontempi$^{1,2}$, 
M. Giroletti$^{1}$\thanks{E-mail: giroletti@ira.inaf.it}, 
F. Panessa$^{3}$ 
M. Orienti$^{1,2}$ 
A. Doi$^{4}$\\
$^{1}$INAF Istituto di Radioastronomia, via Gobetti 101, 40129 Bologna, Italy\\
$^{2}$Dipartimento di Astronomia, Universit\`a di Bologna, via Ranzani 1, 40127 Bologna, Italy\\
$^{3}$INAF/IAPS via del Fosso del Cavaliere 100, 00133 Roma, Italy\\
$^{4}$Institute of Space and Astronautical Science, JAXA, 3-1-1 Yoshinodai, Sagamihara, Kanagawa 229-8510, Japan}
\begin{document}

\date{Date: ?}

\pagerange{\pageref{firstpage}--\pageref{lastpage}} \pubyear{2002}

\maketitle

\label{firstpage}

\begin{abstract}
We report on sensitive dual-frequency (1.7 and 5 GHz) European VLBI Network observations of the central region of nine Seyfert galaxies. These sources are among the faintest and least luminous members of a complete sample of nearby ($d<22$ Mpc) low luminosity AGNs. We detect radio emission on milliarcsecond scale in the nuclei of 4 galaxies, while for the other five sources we set an upper limit of $<\sim100\ \mu$Jy. In three sources, namely NGC\,3227, NGC\,3982, and NGC\,4138, radio emission is detected at both 1.7 and 5 GHz and it is resolved in two or more components. We describe the structural and spectral properties of these features; we find that in each of these three nuclei there is one component with high brightness temperature (typically $T_B >10^{7.5}$~K) and flat/intermediate spectral index ($0.3\le \alpha \le 0.6$, $S(\nu)\sim\nu^{-\alpha}$), accompanied by secondary steep spectrum extended components. In these cases, non-thermal emission from jets or outflows is thus the most natural explanation. A faint feature is detected in NGC\,4477 at 5 GHz; keeping in mind the modest significance of this detection ($\sim 5\sigma$), we propose the hot corona as the origin of non-thermal emission, on the basis of the unrealistic magnetic field values required by synchrotron self-absorption. Finally, the five non-detected nuclei remain elusive and further observations on intermediate scales will be necessary to investigate their nature.
\end{abstract}


\section{Introduction}

Active galactic nuclei are traditionally divided into radio quiet (RQ) and radio loud (RL) depending on the ratio between optical and radio flux density \citep{Kellerman1989}. The nuclear regions of RL AGNs are characterized by high brightness temperature cores and jets with compact knots, in some cases moving with superluminal velocity; on large scale, the radio emission of RL AGNs can reach out to several hundred kiloparsecs, well beyond the size of the host galaxy. RQ AGNs like Seyfert galaxies are much fainter in the radio band and their emission is confined in the sub kpc scale. However, it has become clear in the last years, through VLA surveys, that RQ AGNs are not completely silent in the radio band \citep[e.g.][]{Nagar2002,Ho2001}. While the origin of the radio emission is clear in RL AGNs as synchrotron radiation from energetic particles in jets and lobes, the case of RQ AGNs is not well established. Since the radio structure in the nuclear region is complex, it is of fundamental importance to resolve them with the highest spatial resolution achievable and to obtain spectral information through multi-wavelength radio observations. In some bright targets, VLBI studies have successfully shed light on the properties of these regions, e.g.\ revealing thermal free-free emission from an X-ray heated corona in NGC\,1068 \citep{Gallimore2004}, or two-sided jet-like structures with low speeds, indicating non-relativistic jet motion, possibly due to thermal plasma like in NGC\,4151 \citep{Ulvestad2005} or at most mildly relativistic motion of non-thermal plasma as in NGC\,4278 \citep{Giroletti2005}. 

However, the bulk of the low luminosity radio quiet AGN population is characterized by low radio flux densities which require high sensitivity for a proper study. We thus considered the complete and distance limited sample of 28 local ($d\le22$ Mpc) Seyfert galaxies selected by \citet{Cappi2006}\footnote{The original paper only counted 27 sources, since NGC\,3982 lacked {\it XMM-Newton} data at that time}. In \citet{Giroletti2009}, we observed the cores of five weak ($\sim$ mJy level) targets with the European VLBI Network (EVN) to complement observations of the brighter members of the sample available in the literature \citep[e.g.][]{Trotter1998,Gallimore2004,Ulvestad2005} . Of these, 4 sources were detected, revealing a complex scenario, where diverse underlying physical mechanisms can be responsible for the nuclear radio emission in the four detected targets. The radio spectral indices $\alpha$ (defined such that $S(\nu)\sim\nu^{-\alpha}$) range from steep ($\alpha>0.7$) to slightly inverted ($\alpha=-0.1$), brightness  temperature vary from $T_B=10^5$ K to larger than $10^7$ K and cores are either resolved or unresolved, in one case (NGC\,4051) accompanied by a lobe-like structure  \citep{Giroletti2009}. 

In this paper, we continue the study of the sources in the above mentioned sample discussing dual-frequency EVN observations for 9 more sources never observed on milliarcsecond scale. With this work, we complete the VLBI observations for all the sources in the sample that have a detection on VLA scales in at least one radio frequency. In a companion paper (Panessa et al.\ in prep.), we attempt a statistical study of the multi-wavelength properties of the whole sample. 

This paper is structured as follows:  in Sect.~\ref{s.observations} we describe our observations and the data reduction procedures;  the results are presented in Sect.~\ref{s.results} and their discussion is given in Sect.~\ref{s.discussion}; finally, we summarize the main conclusions in Sect.~\ref{s.conclusions}.

\section{Observations and data reduction}\label{s.observations}

We observed the nuclei of eight Seyfert galaxies with the EVN at 1.7 and 5 GHz in June 2009. The targets are NGC\,3185, NGC\,3227, NGC\,3941, NGC\,3982, NGC\,4138, NGC\,4477, NGC\,4639 and NGC\,4698. Moreover, we present also 1.7 and 5 GHz EVN data for NGC\,5194 obtained in February 2008 from our previous observational campaign and never published before.  All the observations were performed in phase reference mode using eight sub-bands separated by 16 MHz each for an aggregate bit rate of 1\,Gbps, which grants great sensitivity. For sources with suitable declinations, we also used the large collecting area of the Arecibo 300m telescope, in addition to the standard stations of Effelsberg, Jodrell Bank, Medicina, Noto, Onsala, Torun, Shanghai, Urumqi, Westerbork, and Yebes (at 5 GHz only); for details, see Table~\ref{t.log}. Each source was observed for about 3 hours switching between targets in order to improve the coverage of the $(u,v)$-plane. 

Amplitude calibration was done a priori in the standard {\tt AIPS}-based EVN pipeline; phase corrections were obtained using observations of bright nearby calibrators. 
After applying the solutions from the calibrators to all sources, we split the data keeping the 8 IF separated and imported the single source files in {\tt Difmap}. For targets which appeared to be clearly revealed in the image plane, we then went back to {\tt AIPS} and re-split the data averaging the IF, in order to improve the signal-to-noise ratio in the visibilities. This allowed us to self calibrate the target phases for NGC\,3227, NGC\,3982, and NGC\,4138, using a solution interval between 1.5 and 3 minutes to maximize the ratio of good to failed solutions (typically about 5:1)

For the sources not readily detected in the image plane, we still kept the IFs separated to avoid bandwidth smearing. We then searched an area of $2^{\prime\prime} \times 2^{\prime\prime}$ around the phase centre for possibly significant excess in the image plane, both with uniform and natural weights.
For sources where compact components were detected, we also searched for low surface brightness extended emission regions by producing images with natural weights. 

\section{Results}\label{s.results}

Radio emission is detected on milliarcsecond scale at both 1.7 and 5 GHz in the central region of NGC\,3227, NGC\,3982, and NGC\,4138. NGC\,4477 is detected at 5 GHz only with a low signal to noise ratio of $\sim$ 5. For the other sources we made an accurate search in a $2^{\prime\prime} \times 2^{\prime\prime}$ field centered on the observation coordinates determined from low resolution literature images \citep[typically from][]{Ho2001}. NGC\,3185, NGC\,3941, NGC\,4639, NGC\,4698, and NGC\,5194 remain undetected at either frequencies at a level varying between 20 and 160 $\mu$Jy~beam$^{-1}$ ($3\sigma$ rms). The different rms values depend  on the observing frequency and on the presence of sensitive stations (observing runs with Arecibo are much deeper); details about the individual noise value for each source at the two frequencies are given in Table~\ref{t.nondetections}. We only note that (1) some large scale emission is detected in NGC\,5194, but too diffuse and faint to properly image or modelfit and (2) some excess in individual sources is occasionally found at the 5-6 $\sigma$ significance level in 5 GHz images, although it is almost impossible to discriminate whether it is real or simply a statistical fluctuations; we just report the positions of these local peaks in the last column of Table~\ref{t.nondetections} for future reference but do not make any further speculation about their nature in this work.

\begin{figure*}
\centering
\includegraphics[width=0.45\textwidth]{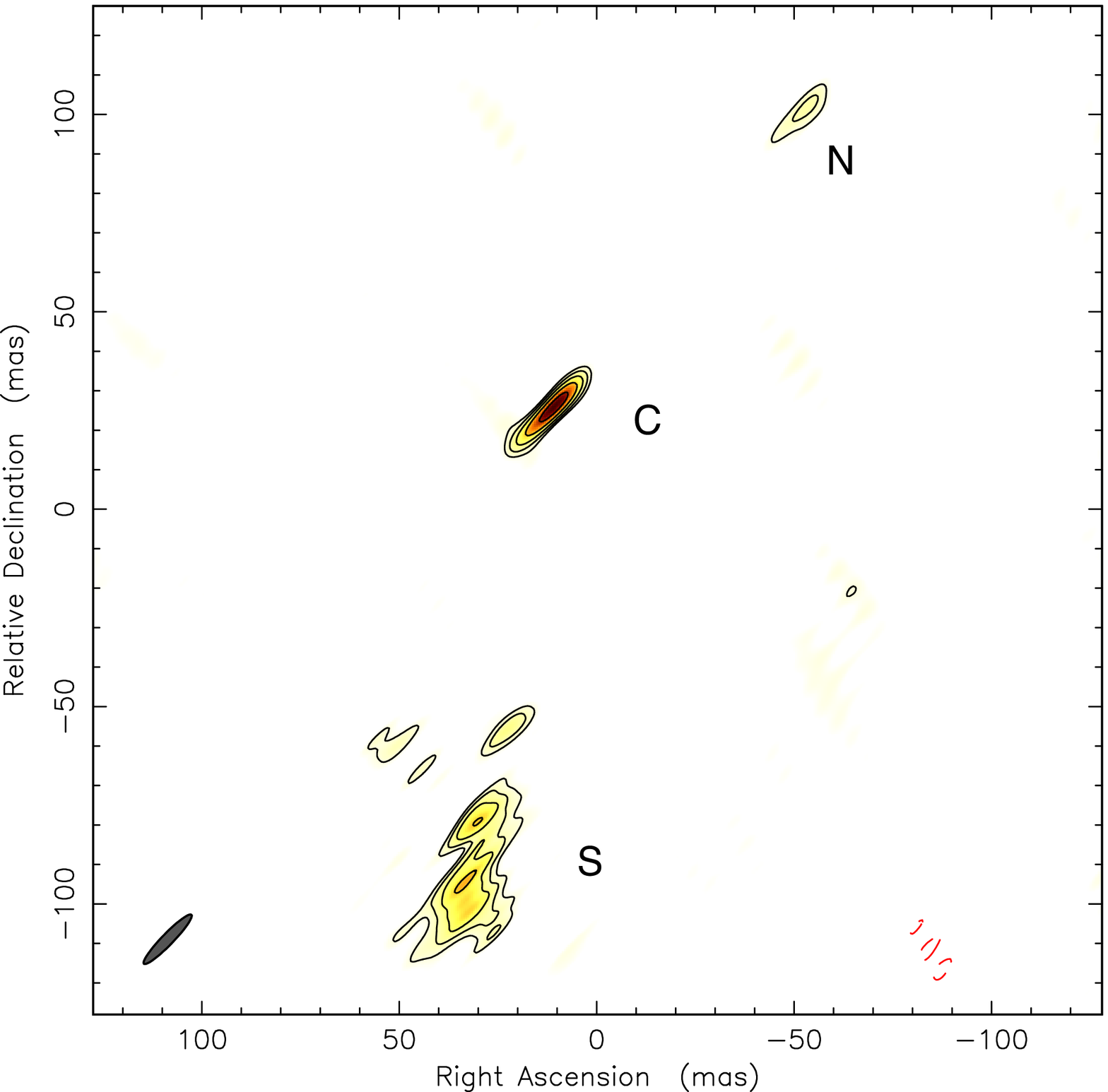}
\includegraphics[width=0.45\textwidth]{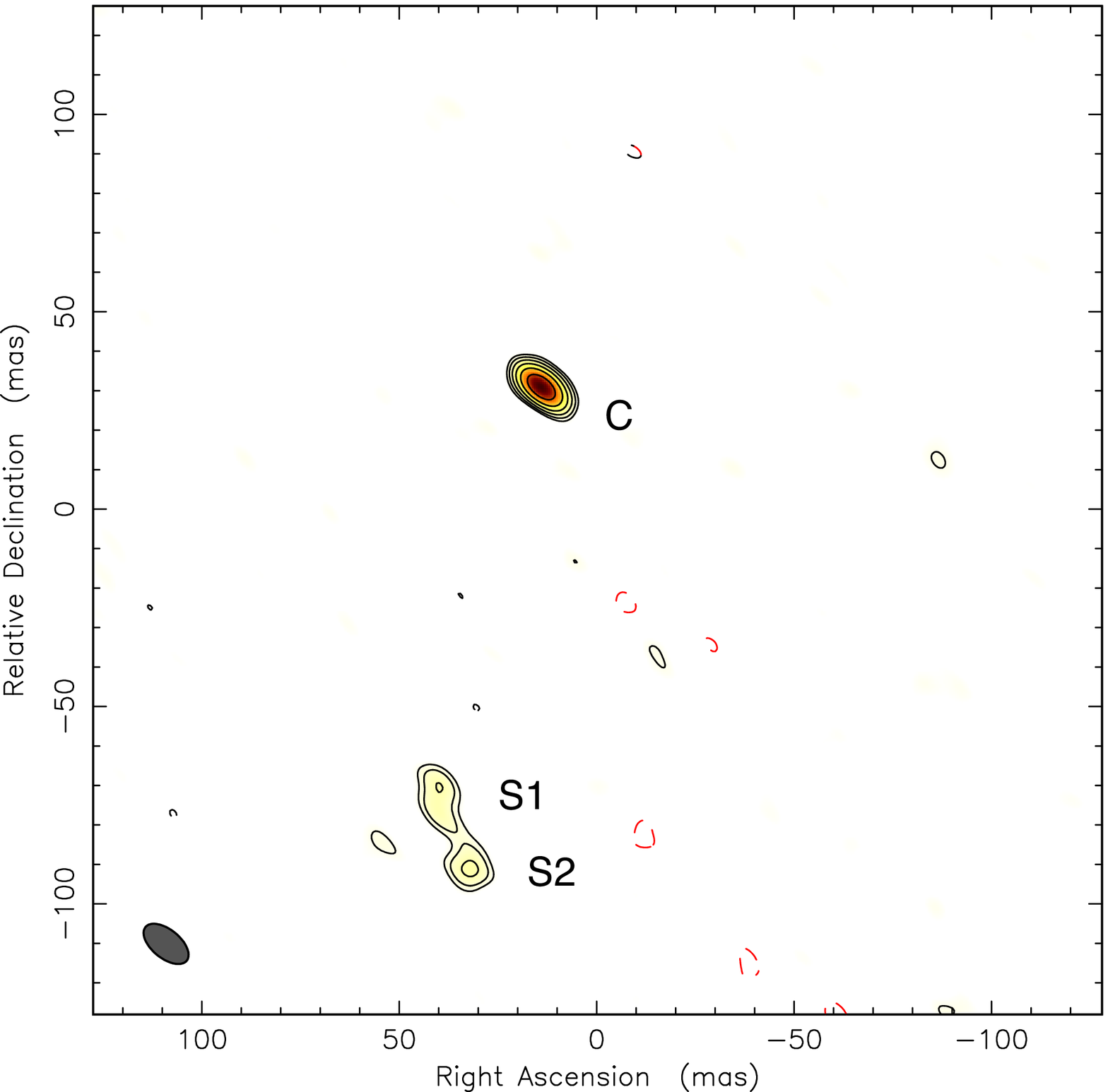}
\caption{Images of NGC\,3227 at 1.7 GHz (left) and 5 GHz (right). Contours are traced at $(-1, 1, 2, 4, \dots) \times$ the $\sim 3\sigma$ noise level, which is 0.13 and 0.08 mJy~beam$^{-1}$ at 1.7 and 5 GHz, respectively. Half-Peak Beam Width (HPBW) are shown in the lower left corner, and their size is 2.9 mas $\times$ 17.3 mas in P.A.\ $-44$\degr\ and 7.2 mas $\times$ 13.5 mas in P.A.\ 50\degr\ at 1.7 and 5 GHz, respectively.}
\label{f.ngc3227}
\end{figure*}

\begin{figure*}
\centering
\includegraphics[width=0.45\textwidth]{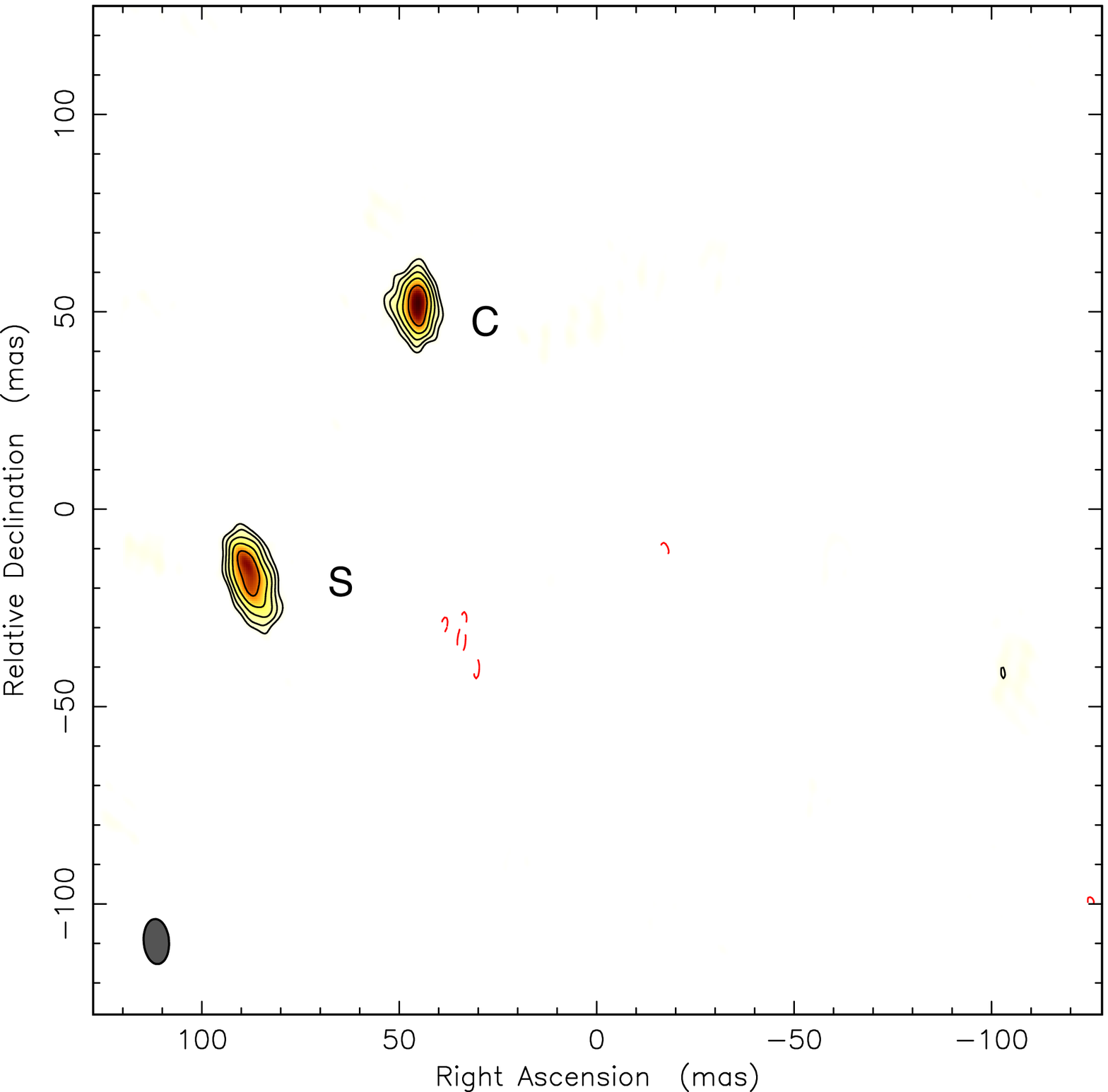}
\includegraphics[width=0.45\textwidth]{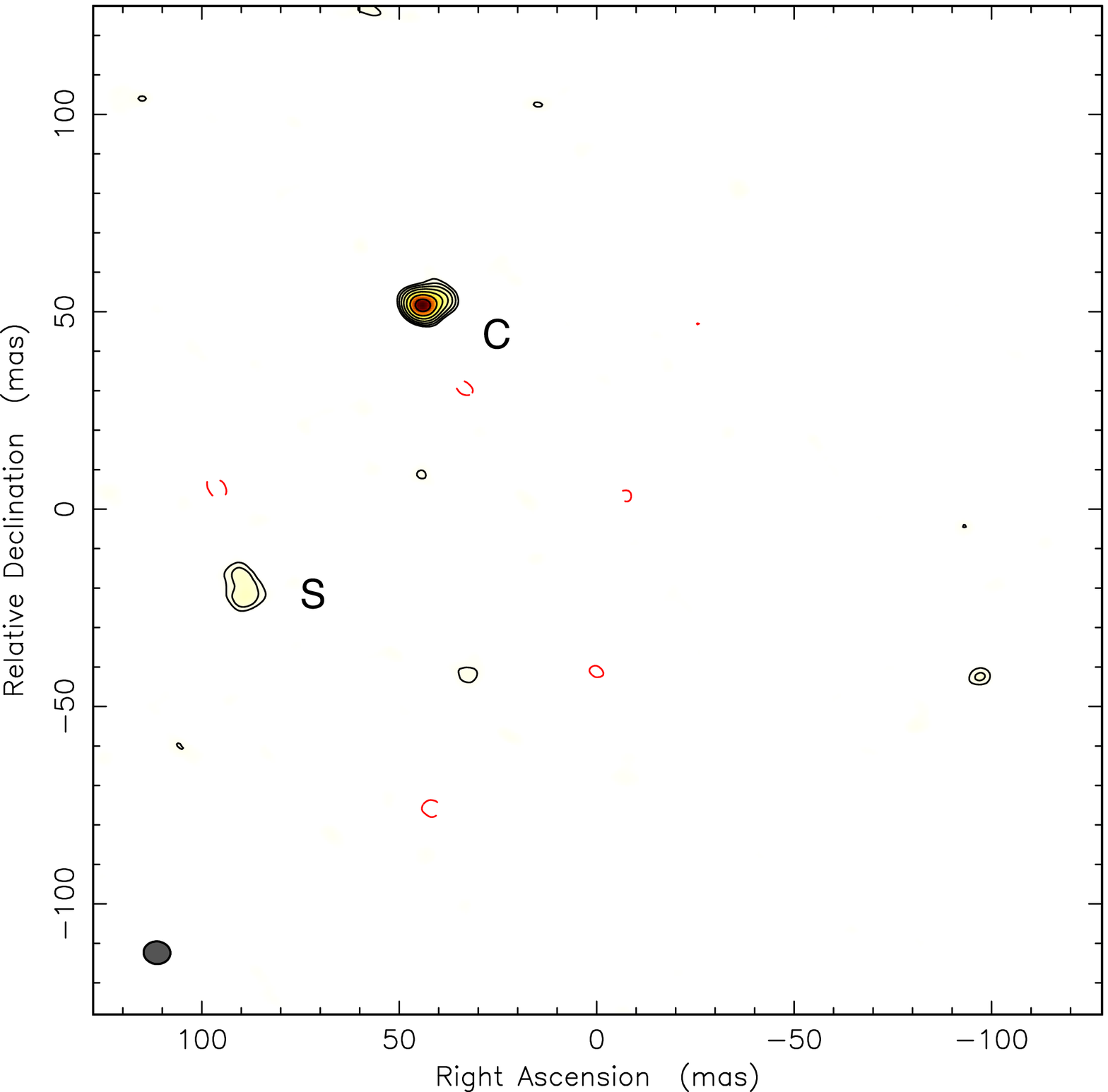}
\caption{NGC\,3982 at 1.7 GHz (left) and 5 GHz (right). Contours are traced at $(-1, 1, 2, 4, \dots) \times$ the $\sim 3\sigma$ noise level, which is 0.20 and 0.09 mJy~beam$^{-1}$ at 1.7 and 5 GHz, respectively. HPBW are shown in the lower left corner, and their size is 6.4 mas $ \times $ 11.4 mas in P.A.\ 4\degr\ and 5.7 mas $ \times $ 6.8 mas in P.A.\ 85\degr\ at 1.7 and 5 GHz, respectively.}
\label{f.ngc3982}
\end{figure*}

\begin{figure*}
\centering
\includegraphics[width=0.45\textwidth]{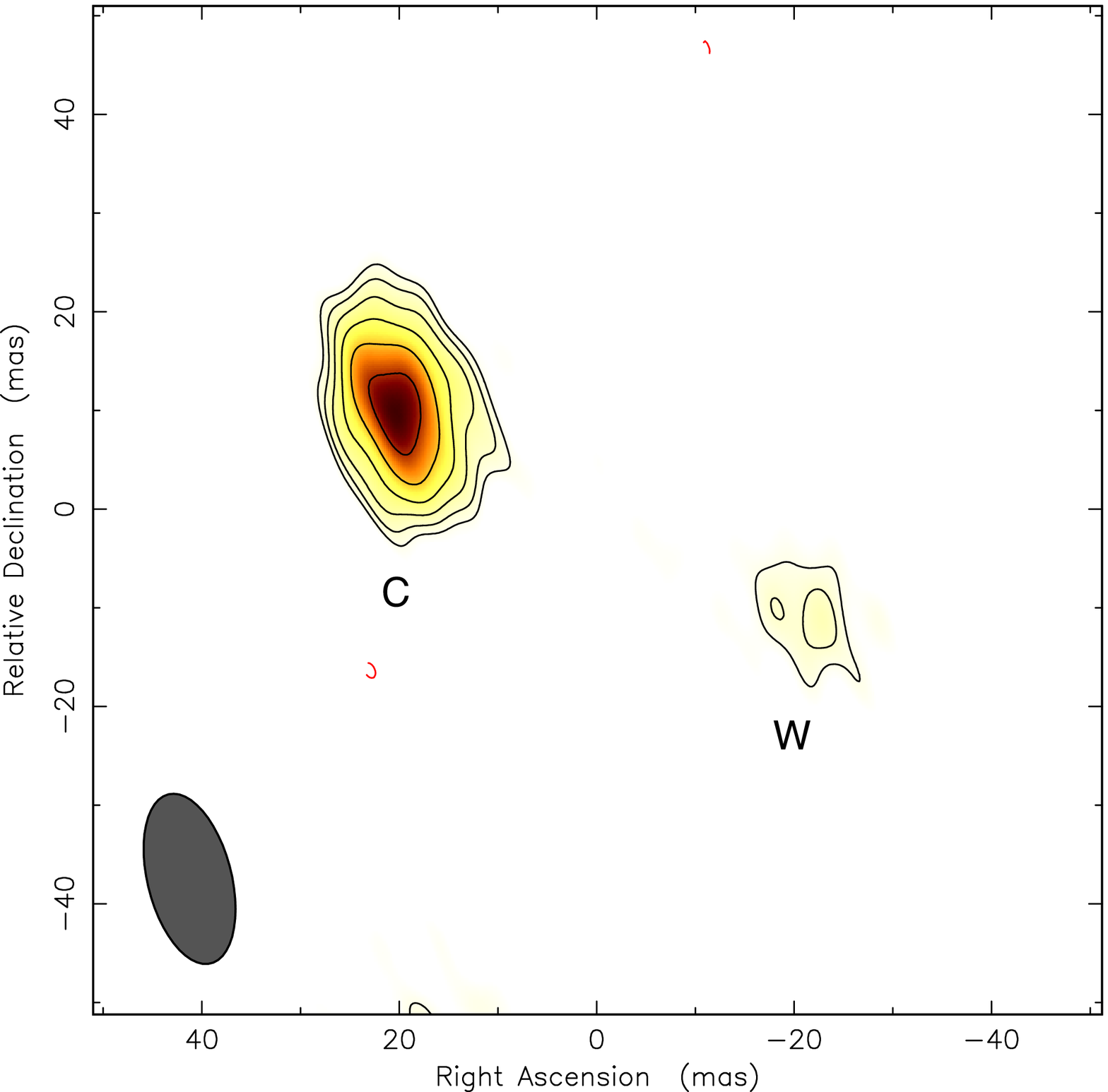}
\includegraphics[width=0.45\textwidth]{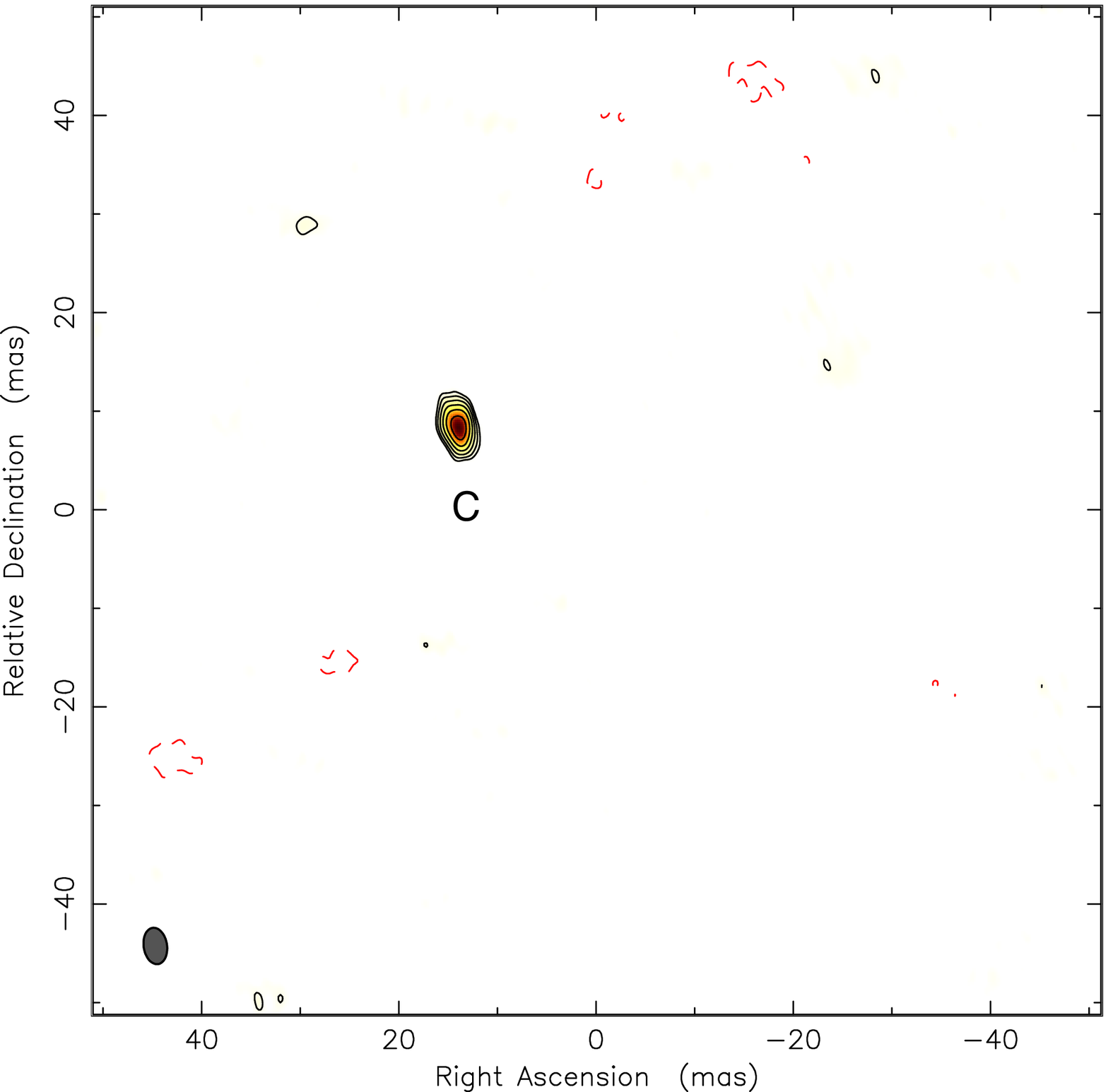}
\caption{NGC\,4138 at 1.7 GHz (left) and 5 GHz (right). Contours are traced at $(-1, 1, 2, 4, \dots) \times$ the $\sim 3\sigma$ noise level, which is 0.14 and 0.09 mJy~beam$^{-1}$ at 1.7 and 5 GHz, respectively. HPBW are shown in the lower left corner, and their size is 8.5 mas $ \times $ 17.7 mas in P.A.\ 14\degr\ and 2.4 mas $ \times $ 3.7 mas in P.A.\ 8\degr\ at 1.7 and 5 GHz, respectively.}
\label{f.ngc4138}
\end{figure*}

The detected galaxies have a total flux density on milliarcsecond scale that ranges from $\sim 9$ mJy (NGC\,3227 at 1.7 GHz) down to about the threshold set by the sensitivity limit of our observations (a few hundreds $\mathrm{\mu Jy}$). In general this corresponds to a broad range of the compact-to-diffuse flux density ratio, with EVN flux densities between 5\% and 100\% of the corresponding VLA flux density in the same sources. We list in Table~\ref{t.cores} the astrometric coordinates and the total flux density with EVN (this work) and the VLA \citep{Ho2001} of the detected sources.
While NGC\,4477 is detected with limited significance and at one frequency only, NGC\,3227, NGC\,3982, and NGC\,4138 are detected at both frequencies at $\gg10\sigma$ confidence level, with one or more components clearly revealed in both bands (see Figs.~\ref{f.ngc3227}, \ref{f.ngc3982}, and \ref{f.ngc4138}). Each component has a size of a few milliarcseconds while the radius of the area over which the features are distributed can reach up to $>\sim200$~mas, which corresponds to linear sizes of several tens of parsecs.

We determined observational parameters for the various features through visibility data model fitting in Difmap (see Table~\ref{t.modelfits}). We used elliptical or circular Gaussian components to describe the visibility data, starting in general from the 1.7 GHz dataset; we then used the obtained model as a starting guess for the 5 GHz data, letting the various parameters free to vary. New components were then added if significant residuals appeared, while we removed components which got unrealistic parameters (e.g. negative flux density). An accurate description of the observational and physical properties of each source is given in the following sub-sections.

\subsection{NGC\,3227}

The radio emission in the nuclear region of the Seyfert 1.5 galaxy NGC\,3227 is resolved in a few subcomponents; we show the 1.7 and 5 GHz EVN images in Fig.~\ref{f.ngc3227}. The model-fit parameters of the various features are reported in Table~\ref{t.modelfits}. The main feature is a compact component, detected at both 1.7 and 5 GHz, accompanied by some diffuse emission. We label with C the most compact component, with S the southern one (resolved in S1 and S2 at 5 GHz), and with N the northern one (present only at 1.7 GHz). 

Component C is found in position R.A.\ 10$^h$ 23$^m$ 30.573$^s$, Dec.\ 19$^\circ$ 51\arcmin 54.274\arcsec and it is basically unresolved, with a size $\theta_{1.7}=5.9$ mas and $\theta_5=1.2$ mas at the two frequencies. With a flux density of $S_{1.7}=1.2$ mJy and $S_{5}=0.6$ mJy, component C is characterized by a relatively flat spectrum $\alpha \sim 0.6$. The more extended southern emission region S is located about $\sim 120$ mas ($\sim12$ pc, assuming $D=20.6$ Mpc) south of C in position angle $+170^{\circ}$ and it has a size of $\sim 50$ mas; this region is best described by a single elliptical Gaussian at 1.7 GHz with $S_{1.7}=6.7$ mJy and by two nearby circular Gaussians S1 and S2 at 5 GHz, with a summed flux density of $S_5=0.52$ mJy. Even accounting for possible uncertainties on the absolute amplitude scale and for significant resolution effects, it is clear that the southern extended emission has a much steeper spectral index ($\alpha \sim 2.3$) than the main one. Finally, some weak and diffuse emission is found also north of C, at about 90 mas in P.A. $45^{\circ}$. This emission (component N) is only detected at 1.7 GHz, consistent with having a rather steep spectrum, too.

The VLBI flux density measured in our observations is about 15\% of the VLA peak flux density at 1.7 GHz and about 7\% at 5 GHz (Ho \& Ulvestad 2001).

\subsection{NGC\,3982}

NGC\,3982 is a Seyfert 1.9 galaxy whose central region is also characterized by the presence of at least two milliarcsecond scale components detected at both 1.7 and 5 GHz, as shown in Fig.~\ref{f.ngc3982}.  We label the two components as N and S, for the northern and southern feature, respectively. Similarly to components C and S in NGC\,3227, the two features are separated by $\sim 84$ mas in P.A.\ $+148^{\circ}$ ($\sim 8.5$ pc, assuming $D=20.5$ Mpc) and present distinct spectral properties. Component N has a flux density of $S_{1.7}=1.5$ mJy and $S_{5}=0.9$ mJy, corresponding to $\alpha\sim0.4$, while S has a much steeper spectrum, with  $S_{1.7}=1.8$ mJy, $S_{5}=0.4$ mJy, and $\alpha\sim1.4$. Both components are rather extended but higher signal-to-noise ratio on the long baselines would be required to resolve the details of the possible substructures. A third, weaker component is present in the residual images at 5 GHz about 200 milliarcseconds west of S; since a mildly significant excess is also present at 1.7 GHz, this may also be a real feature, although future more sensitive observations will be required to confirm it.

The total VLBI flux density in our 5 GHz VLBI image is $\sim65\%$ of the VLA flux density, while it accounts for 100\% (within the uncertainty) at 1.7 GHz \citep{Ho2001}.

\subsection{NGC\,4138}

NGC\,4138 is a Seyfert 1.9 galaxy; it presents radio emission in its nuclear region, with a structure resolved in two components labelled as C and W (detected only at 1.7 GHz) in Fig.~\ref{f.ngc4138}. The eastern component (C) is brighter ($S_{1.7}=1.0$ mJy and $S_{5}=0.75$ mJy from visibility model fits) and has a moderately flat spectrum ($\alpha\sim0.3$). The position of the component is not entirely consistent at the two frequencies, with the 1.7 GHz position being offset by 6 mas eastward with respect to the 5 GHz dataset. The western component is only detected at 1.7 GHz, with a flux density of 0.3 mJy and separated by $\sim 50$ mas (3.5 pc at $D=13.80$ Mpc) from C.

The milliarcsecond emission revealed in our maps accounts for all the flux density detected by the VLA at both frequencies, suggesting that no significant structure is present on larger scale.

\subsection{NGC\,4477}

NGC\,4477 is a Seyfert 2 galaxy. At 5 GHz, we detect a $\sim 5\sigma$ excess in position R.A.\ 12$^{h}$ 30$^{m}$ 02.203$^{s}$,  Dec.\ 13$^\circ$ 38\arcmin 12.856\arcsec. It is difficult to make strong claims about the reliability of this detection. However, differently from the very uncertain local peaks reported for the other sources in Table~\ref{t.nondetections}, the analysis of the NGC\,4477 dataset makes us more confident about this detection:\ we have tried to split the data in different time ranges, different Stokes parameters (LL and RR), and different sub-bands. A few $\sigma$ excess remains present in each of the subset of the data. Of course, only repeated observations will make it possible to confirm the detection. 

In any case, the flux density of the component would be around 0.14 mJy, which accounts for $\sim100\%$ of the VLA flux density reported by \citet{Ho2001}; the modelfit procedure yields a size of about 1.5 mas ($\sim 0.12$ pc, assuming $d=16.8$ Mpc). At 1.7 GHz, no significant excess is found in the image, which has a very low image noise of about $1\sigma \sim 8 \mu$Jy~beam$^{-1}$. Assuming a $5\sigma$ upper limit and a compact source, we estimate an inverted spectral index of $\alpha \sim -1.2$, although the associated uncertainty is certainly very large.

\section{Discussion}\label{s.discussion}

The analysis of the structural and spectral properties of the radio emission in the nuclei of Seyfert galaxies is a key tool to constrain the physical processes at work in these regions. Past studies \citep[e.g.][and many others]{Trotter1998,Falcke2000,Nagar2002,Gallimore2004,Giroletti2009} as well as the results obtained in the present work reveal a complex picture with both compact features and resolved structures, inverted/flat to intermediate/steep spectra, and radio luminosity ranging over several orders of magnitude. Moreover, a significant fraction of nuclei remains undetected even down to very low flux density levels provided by the current sensitive observations.

In a simplified scheme, steep spectrum regions are interpreted as produced via synchrotron radiation by a population of relativistic electrons and are typically associated to relatively diffuse emission; flat or inverted spectral indices are rather ascribed to a synchrotron self absorption mechanism (SSA, yet produced by a population of relativistic particles) or to a thermal process like Bremsstrahlung. In this context, components S and N in NGC\,3227, component S in NGC\,3982, and component W in NGC\,4138 are rather straightforward to interpret as regions of synchrotron radiation from relativistic particles, although it is not trivial to constrain the origin of the particles and the acceleration mechanism at work in these regions.

The remaining components have somewhat flatter spectra, ranging from $\alpha\sim0.6$ (C in NGC\,3227) to $\alpha \sim0.3$ (C in NGC\,4138), up to the case of the tentative detection in NGC\,4477 where the spectral index would be inverted. In this cases, a useful diagnostic to discriminate between SSA and thermal processes is the brightness temperature, defined as follows:

\begin{equation}
T_B=\frac{S(\nu)}{2k\theta_{\rm maj} \theta_{\rm min}}\left(\frac{c}{\nu}\right)^2
\end{equation}

\noindent
where $S(\nu)$ is the flux density at the frequency $\nu$, $\theta_{\rm maj}$ e $\theta_{\rm min}$ are major and minor axis respectively, $k$ is the Boltzmann constant and $c$ the speed of light. High values of $T_B$ ($\gg10^6$ K) can arise only from a non thermal process and are interpreted as signatures of AGN-like, non thermal radio emission, most likely powered by underfed black holes \citep[e.g.][]{Falcke2000}. 

In Tables~\ref{t.physics1} and~\ref{t.physics5}, we list the brightness temperatures and other physical parameters of the various components found in our images at 1.7 and 5 GHz, respectively. In the sources NGC\,3227, NGC\,3982, and NGC\,4138, we find a component with $\log T_B > 7.5$, and reaching up to $\log T_B =9.1$ in the main component of NGC\,3982 considering the 1.7 GHz data. These features are thus interpreted as AGN-like cores, hence they are labelled with C. Other physical parameters are then determined under the assumption that the emission is synchrotron radiation and that the source is a prolate spheroid with volume $V$ given by:

\begin{equation}\label{eq-volume}
V=\frac{\pi}{6}d_{\rm maj}d_{\rm min}^2
\end{equation}

\noindent
where $d_{\rm maj}$ and $d_{\rm min}$ are the linear size of the major and minor axis. We further assume that the volume of the source is completely and homogeneously filled with a relativistic plasma (i.e.\ a filling factor of 1), that proton and electron densities are the same, and that we are in minimum energy conditions. Under these assumptions, we derive the energy density $U_{\rm min}$ and the equipartition magnetic field $B_{\rm eq}$ from the following equations \citep[e.g.][]{Orienti2012}:

\begin{equation}\label{eq-energy}
U_{\rm min}=   7\times 10^{24}  \left( \frac{L}{V} \right)^{4/7}
\end{equation}

\begin{equation}\label{eq-magneticfield}
B_{\rm eq}=  \sqrt{\frac{24}{7} \pi u_{\rm min}}
\end{equation}

We report the derived quantities in Tables~\ref{t.physics1} and~\ref{t.physics5}; the values are in good agreement with the expectations for compact regions. In general, the equipartition magnetic fields $B_{\rm eq}$ are of the order of a few mG; larger values are found for the most compact C features. The differences in the values obtained at the two frequencies provide an estimate of the associated uncertainty, which is an unavoidable consequence of the many assumptions and of the intrinsic errors in the observations and model fit procedures.

Overall, it emerges a picture in which we have highlighted the compact base of a jet or outflow in at least three sources; these jets or outflows are also directly revealed thanks to the observation of the secondary, more extended steep spectrum features such as NGC\,3227's S1, S2, and N, NGC\,3982 S, and NGC\,4138 W. In the case of NGC\,3227, our observations directly connect with the structure and physical properties revealed by earlier dual frequency MERLIN observations. Indeed, \citet{Mundell1995} reported a double structure elongated in PA $-10^{\circ}$, with a steep ($\alpha=0.9$) spectral index. Our VLBI observations resolve the southern MERLIN component in the three substructures C, S, and N. C and S are also aligned with the same PA of the more extended $\sim 1^\prime$ emission, while N is consistent with a lateral extension visible in the main MERLIN 5 GHz component. The optical position of the nucleus is also consistent with the main MERLIN component, so it is a further argument in favour of component C as the actual core, at the base of a two sided outflow of non-thermal material. Finally, albeit with the significant uncertainty described above and taking into account the different spatial resolution of the two instruments, the equipartition magnetic fields from EVN and MERLIN ($B\sim2$ mG) turn out to be in reasonable agreement. 

As to the nature of the ejected material, while the typical signatures of the synchrotron emission reveal that it is non-thermal, it is much less clear whether it also has a bulk relativistic motion typical of jets in RL AGNs or if it has much slower velocity. In the absence of clear indicators of beaming, the latter possibility seems more likely. \citet{Orienti2010} have revealed a similar scenario in nearby Seyfert galaxies and find evidence for non-thermal radiation related to AGNs, perhaps from a jet that is disrupted because of a dense environment.

In the remaining sources, the scenario is less well constrained. For the only other detected source, NGC\,4477, we have an intermediate brightness temperature $\log T_B = 6.5$ and a very inverted spectrum, compatible either with SSA or a thermal process. In this case, we use our data to try to estimate other physical parameters, such as the magnetic field necessary to have the SSA spectral turnover in a range compatible with what we observe. The typical magnetic field is expected to be of the order of a few mG. Under the assumption that the observed spectrum is shaped by SSA, the magnetic field $H$ can be estimated as follows:

\begin{equation}
H=\frac{\nu_p^5\theta^{4}}{f(\alpha)^5 S_p(\nu)^2(1+z)}
\label{campo-mag-sinc-auto}
\end{equation}

\noindent
where $f(\alpha)$ is a function that weakly depends on the spectral index \citep[for $\alpha \sim 0.5$ its value  is about 8, see][]{Kellerman1981}.

Assuming a turnover frequency $\nu_p\sim5$\,GHz and the flux and size derived from the modelfit, we obtain a value of $H \sim 10^9$ G, a so high value that we can exclude the synchrotron self absorption as the physical process at the origin of the emission and call instead a thermal process. On the other hand, we can estimate the electron density $n_e$:

\begin{equation}
\label{densita elettroni}
n_e=\sqrt{1.84\times 10^{41}  \left(\frac{T}{10^4}\right)D_L^2S_{\nu}V^{-1}g_{ff}^{-1}}
\end{equation}

\noindent
where $S(\nu)$ is the flux density at the frequency $\nu$, $T$ the temperature in unit of $10^4$~K,  $g_{ff}$ the Gaunt factor that at the radio frequency is equal to $g_{ff}\sim17.7+\ln{(T^{3/2}/\nu)}$. The resulting electron density for NGC\,4477 would then be $n_e=7.2\times10^4$ m$^{-3}$, compatible with a thermal process. It is thus likely that the emission is produced in a compact (some 0.1's pc) region at the heart of the AGN, such as the hot corona that surrounds the accretion disk. A similar scenario was proposed in the observation of NGC\,1068 by \citet{Gallimore2004}, who revealed an inverted spectral index between 1.7 and 5 GHz and flat above 5 GHz in the innermost component; they invoke a thermal process to explain the emission and identify the region responsible of the emission with the hot corona. \citet{Laor2008} suggested that, by analogy with coronally active cool stars, the expected $L_{\rm 5 GHz}/L_{\rm 0.2-20 keV}$ ratio should be around 10$^{-5}$, in the case where the X-ray and radio emission are both produced within the hot corona. For NGC\,4477, we extrapolated the 0.2-20 keV X-ray luminosity from the 2-10 keV luminosity assuming a photon index of 1.9 \citep[see][]{Cappi2006} and derive $L_{\rm 5 GHz}/L_{\rm 0.2-20 keV} \sim 10^{-3}$, suggesting that some extra radio component (e.g., nuclear HII regions) may contribute to the radio total luminosity.

The five remaining sources, NGC\,3185, NGC\,3941, NGC\,4639, NGC\,4698, and NGC\,5194 do not possess compact features at the sensitivity level of our observations. All but NGC\,5194 have radio flux density on arscecond scale around 0.2 mJy at 5 GHz, while they are undetected at 1.7 GHz at the 0.15 mJy\,beam$^{-1}$ level \citep{Ho2001}. The inverted spectral index could again suggest thermal emission, although on an angular scale intermediate between VLBI and VLA. The upgraded e-MERLIN facility might be ideal to study this sources. Indeed, e-MERLIN could provide an excellent combination of sensitivity and angular resolution to study all of the sources in our sample, permitting to highlight the faint jet/outflow structures detected in the VLBI images.

\section{Conclusions}\label{s.conclusions}

The aim of this work was to observe the nuclear region of 9 local Seyfert galaxies in the radio band on the parsec scale. These galaxies have been extracted from the complete sample of 28 Seyfert studied in \citet{Cappi2006} with a VLA detection but still lacking  high resolution (VLBI) observations. At 1.7 GHz we revealed 3 sources (NGC\,3227, NGC\,3982, and NGC\,4138), while at 5 GHz we were able to reveal radio emission also in the core of NGC\,4477, raising the detection rate to $\sim 44\%$. 

In the sources detected at both frequencies, i.e.\ NGC\,3227, NGC\,3982, and NGC\,4138, we detected one component with high brightness temperature ($\log T_B>7.5$) and flat/intermediate spectral index ($0.3\le \alpha \le 0.6$), which we ascribe to non-thermal emission from the immediate vicinity of the central black hole; moreover, steep spectrum extended components are also detected within some tens to hundred milliarcseconds from the core, suggesting the presence of jets or outflows on parsec scales. Indeed, the VLBI structure in NGC\,3227 connects nicely to the larger scale emission observed in literature MERLIN images. The physical parameters estimated under the assumption of minimum energy are quite reasonable; e.g., the equipartition magnetic field are of a few mG, and slightly higher in the most compact features.

In NGC\,4477, which is detected only at 5 GHz, the brightness temperature is somewhat lower, and the physical parameters seem at odd with a synchrotron self-absorption scenario, mainly because of the too high magnetic field required. On the other hand, a thermal free-free origin for its radio emission seems more viable, similar to what found in NGC\,1068 by \citet{Gallimore2004}. However, we remind that the significance of the detection is limited ($\sim5\sigma$) and future observations will be needed to confirm this speculation.

Finally, NGC\,3185, NGC\,3941, NGC\,4639, NGC\,4698, and NGC\,5194 remain undected down to very low brightness levels ($<\sim100\ \mu$Jy beam$^{-1}$). Since all these sources have weak but compact cores in VLA images, observations at intermediate resolution will be necessary to reveal the structure and characterize the physical condition of their nuclear regions.

\section*{Acknowledgements}
   The European VLBI Network is a joint facility of European, Chinese,
   South African and other radio astronomy institutes funded by their
   national research councils.
  This effort/activity is supported by the European Community Framework
  Programme 7, Advanced Radio Astronomy in Europe, grant agreement No. 227290.
We acknowledge a
contribution from the Italian Foreign Affair Minister under the bilateral
scientific collaboration between Italy and Japan.

\newpage 

\clearpage
\newpage

\begin{table*}
\caption{Observation log. Station codes are as follows: Ar: Arecibo, Ef: Effelsberg, Hh: Hartebeesthoek, JB1: Jodrell Bank (Lovell telescope), Jb2: Jodrell Bank (Mk2), Mc: Medicina, Nt: Noto, On: Onsala, Sh: Shanghai, Tr: Torun, Ur: Urumqi, Wb: Westerbork Synthesis Radio Telescope, Ys: Yebes. Stations in parentheses participated in the observations but did not produce good data, either because of technical problems or visibility constraints. }
\footnotesize
\begin{tabular}{llll}
\hline
Date & Freq. & Source name & Participating stations \\
 & (GHz) &   \\
\hline
2008 Feb 28 & 1.7 & NGC\,5194 & Ef, (Hh), Jb1, Mc, Nt, On, Sh, Tr, Ur, Wb \\
2008 Mar 10 & 5 & NGC\,5194 & (Ef), (Hh), Jb1, Mc, Nt, On, Sh, Tr, Ur, Wb \\
2009 Jun 02 & 1.7 & NGC\,3185, NGC\,4477 & Ar, Ef, Jb1, Mc, Nt, On, Tr, (Ur), Wb \\
2009 Jun 08 & 1.7 & NGC\,3227, NGC\,4639 & Ar, Ef, Jb2, Mc, Nt, On, Tr, (Ur), Wb \\
2009 Jun 09 & 1.7 & NGC\,3941, NGC\,3982, NGC\,4138, NGC\,4698 & Ef, Jb2, Mc, Nt, On, Sh, Tr, Ur, Wb \\
2009 Jun 13 & 5 & NGC\,3185, NGC\,4477 & Ar, Ef, Jb2, (Mc), Nt, On, Tr, Ur, Wb, Ys \\
2009 Jun 14 & 5 & NGC\,3227, NGC\,4639 & Ar, Ef, Jb2, (Mc), Nt, On, Tr, Ur, Wb, Ys \\
2009 Jun 15 & 5 & NGC\,3941, NGC\,3982, NGC\,4138, NGC\,4698 & Ef, Jb2, Mc, Nt, On, Sh, Tr, Ur, Wb, (Ys) \\
\hline
\end{tabular}
\label{t.log}
\end{table*}

\begin{table*}
\caption{Summary for non detected sources. We report the peak position in Cols.\ 6 and 7 only for components that are most significant (all at 5 GHz); however, given the large searched area, it is well possible that these excesses are simple statistical fluctuations and we do not consider them to be real detections.}
\footnotesize
\begin{tabular}{lcccccc}
\hline
Source & \multicolumn{2}{c}{Phase tracking position} & 1.7 GHz $3\sigma$ rms  & 5 GHz $3\sigma$ rms & \multicolumn{2}{c}{peak position}   \\
 & (hh mm ss) & ($^\circ$~~\arcmin~~\arcsec) & ($\mu$Jy~beam$^{-1}$) & ($\mu$Jy~beam$^{-1}$) & (hh mm ss) & ($^\circ$~~\arcmin~~\arcsec) \\
(1) & (2) & (3) & (4) & (5) & (6) & (7) \\
\hline
NGC\,3185 & 10:17:38.660 & 21:41:17.400 & 20 & 27 & \ldots & \ldots \\
NGC\,3941 & 11:52:55.363 & 36:59:10.890 & 114 & 88 & 11 52 55.347 & 36 59 11.908 \\
NGC\,4639 & 12 42 52.363 & 13 15 26.750 & 30 & 50 & 12 42 52.381 & 13 15 26.604 \\
NGC\,4698 & 12 48 22.919 & 08 29 14.550 & 160 & 95 & 12 48 22.938 & 08 29 14.623 \\
NGC\,5194 & 13 29 52.804 & 47 11 40.065 & 75 & 160 & \ldots & \ldots\\
\hline
\end{tabular}
\label{t.nondetections}
\end{table*}

\begin{table*}
\caption{Summary for detected sources. We report the astrometric position for the main component in Cols.\ 2 and 3, the total flux density of the source in EVN data in Cols.\ 4 and 5, and the core flux density in VLA data from \citep{Ho2001} in Cols. 6 and 7.}\label{t.cores}
\begin{center}
\begin{tabular}{lcccccc}
\hline
Name & R.A. & Dec. & $S_{\rm 1.7\ GHz,\ EVN}$ & $S_{\rm 5\ GHz,\ EVN}$ & $S_{\rm 1.4\ GHz,\ VLA}$ & $S_{\rm 5\ GHz,\ VLA}$ \\
 & (hh mm ss) & (\degr~\arcmin~\arcsec) & (mJy) & (mJy) & (mJy) & (mJy) \\
 \hline
NGC 3227 & 10 23 30.573 & 19 51 54.274 & 9.0 & 1.12 & 78.2 & 25.9 \\
NGC 3982 & 11 56 28.165 & 55 07 30.917 & 3.2 & 1.3 & 3.56 & 1.79 \\
NGC 4138 & 12 09 29.802 & 43 41 06.875 & 1.3 & 0.74 & 0.45 & 0.78 \\
NGC 4477 & 12 30 02.203 & 13 38 12.856 & \ldots & 0.14 & \ldots & 0.18 \\  
\hline
\end{tabular}
\end{center}
\end{table*}

\begin{table*}
\caption{Parameters of components: (1) galaxy name, (2) name of component, (3) flux density at 5~GHz, (4) relative right ascension at 5~GHz with respect to the C1, (5) relative declination at 5~GHz with respect to the C1, (6) component size and orientation at 5~GHz, (7) flux density at 1.7~GHz, (8) relative right ascension at 1.7~GHz with respect to the C1 at 5~GHz, (9) relative declination at 1.7~GHz with respect to the C1 at 5~GHz, (10) component size and orientation at 1.7~GHz, (11) spectral index.
}\label{t.modelfits}
\begin{tabular}{llccccccccc}
\hline\hline
Name & Comp. & $S_5$ & $\Delta\alpha_5$ & $\Delta\delta_5$ & Size$_5$ & $S_{1.7}$ & $\Delta\alpha_{1.7}$ & $\Delta\delta_{1.7}$ & Size$_{1.7}$ & $\alpha$ \\
 &  & (mJy) & (mas) & (mas) & (mas $\times$ mas, \degr) & (mJy) & (mas) & (mas) & (mas $\times$ mas, \degr) \\
(1) & (2) & (3) & (4) & (5) & (6) & (7) & (8) & (9) & (10) & (11) \\
\hline
NGC 3227 & C & 0.60 & 0 & 0 & $1.2\times1.2, 0$ & 1.22 & -3.0 & -5.2 & $3.8\times 5.9, -31$ & 0.6 \\
 & S1 & 0.27 & 24.7 & -103.5 & $8.3\times8.3, 0$ & \ldots & \ldots & \ldots & \ldots \\
 & S2 & 0.25 & 18.7 & -122.1 & $5.5\times5.5, 0$ & 6.75 & 18.9 & -118.3 & $31.3\times 48.0, 2$ & 2.3(*) \\
 & N & \ldots & \ldots & \ldots & \ldots & 1.07 & -64.8 & 62.8 & $10.3\times 53.2, 18$ \\
NGC 3982 & C & 0.91 & 0 & 0 & $1.3\times 1.3, 0$  & 1.38 & 0.8 & 2.5 & $0.9\times 0.9, 0$ & 0.4 \\
 & S & 0.44 & 45.8 & -71.4 & $61.\times 9.6, 18$ & 1.79 & 43.3 & -67.6 & $8.3\times 8.3, 0$ & 1.3 \\
NGC 4138 & C & 0.74 & 0.0 & 0.0 & $1.7\times 0.6, 25$ & 1.07 & 6.3 & 1.4 & $1.4\times 1.4, 0$ \\
 & W & \ldots & \ldots & \ldots & \ldots & 0.25 & -42.9 & -20.9 & $5.5\times 11.3, 11$ & 0.3 \\
NGC 4477 & C & 0.14 & 0 & 0 & $1.7\times 1.7, 0$ & \ldots & \ldots & \ldots & \ldots \\
\hline
\end{tabular}
\medskip 

(*) The spectral index has been computed between components S at 1.7 GHz and the sum S1+S2 at 5 GHz. 
\end{table*}

\begin{table*}
\caption{Physical quantities at 1.7 GHz.}
\centering
\footnotesize
\begin{tabular}{llccccc}
\hline
Source & Component & $\log T_B$ & $\log L$ & $\log V$ & $\log U_{\rm min}$ & $B_{\rm eq}$ \\
 & & (K) & (W Hz$^{-1}$) & (cm$^{-3}$) & (erg cm$^{-3}$) & (mG) \\
\hline
NGC\,3227 & C & 7.5 & 19.8 & 54.1 & $-$5.76 & 4.3 \\
& S & 6.5 & 20.5 & 56.9 & $-$6.90 & 1.2 \\
& N & 6.1 & 19.7 & 55.9 & $-$6.83 & 1.3 \\
NGC\,3982 & C & 9.1 & 19.8 & 52.0 & $-$4.53 & 17.8 \\
& S & 7.2 & 20.0 & 54.9 & $-$6.14 & 2.8 \\
NGC\,4138 & C & 8.5 & 19.4 & 52.1 & $-$4.86 & 12.2 \\
 & W & 6.6 & 18.8 & 54.8 & $-$6.73 & 1.4 \\
\hline
\end{tabular}
\label{t.physics1}
\end{table*}

\begin{table*}
\caption{Physical quantities at 5 GHz.}
\centering
\footnotesize
\begin{tabular}{llccccc}
\hline
Source & Component & $\log T_B$ & $\log L$ & $\log V$ & $\log U_{\rm min}$ & $B_{\rm eq}$ \\
 & & (K) & (W Hz$^{-1}$) & (cm$^{-3}$) & (erg cm$^{-3}$) & (mG) \\
\hline
NGC\,3227 & C & 7.5 & 19.5 & 52.4 & $-$4.96 & 10.9 \\
& S1 & 5.4 & 19.1 & 54.9 & $-$6.61 & 1.6 \\
& S2 & 5.8 & 19.1 & 54.4 & $-$6.31 & 2.3 \\
NGC\,3982 & C & 7.6 & 19.7 & 52.5 & $-$4.93 & 11.2 \\
& S & 5.7 & 19.3 & 54.7 & $-$6.37 & 2.1 \\
NGC\,4138 & C & 7.8 & 19.2 & 51.4 & $-$4.52 & 18.1 \\
NGC\,4477 & C & 6.5 & 18.7 & 52.7 & $-$5.59 & 5.2 \\
\hline
\end{tabular}
\label{t.physics5}
\end{table*}

\end{document}